\documentclass[aps, prd, reprint, superscriptaddress, amsmath, amssymb]{revtex4-2}

\usepackage [pdftex] {graphicx}
\usepackage {epstopdf}

\usepackage{dcolumn}
\usepackage{bm}
\usepackage[utf8]{inputenc}
\usepackage{color,xcolor}

\usepackage{subcaption}

\usepackage{url}
\usepackage{hyperref}
\usepackage{booktabs}
\usepackage{amsmath, graphicx,algorithm,algorithmic,amsfonts,float,multirow}

\usepackage{diagbox}

\begin{document}
	
\preprint{APS/123-QED}

\title{Conditional Autoencoder for Generating Binary Neutron Star Waveforms with Tidal and Precession Effects}

\author{Mengfei Sun}
\affiliation{%
    Department of Physics,Chongqing University, 
Chongqing 401331, P.R. China\\
}%
 \affiliation{%
    Chongqing Key Laboratory for Strongly Coupled Physics, Chongqing University, Chongqing 401331, P.R. China\\
}%

\author{Jie Wu}%
\affiliation{%
    Department of Physics,Chongqing University, 
Chongqing 401331, P.R. China\\
}%
\affiliation{%
Chongqing Key Laboratory for Strongly Coupled Physics, Chongqing University, Chongqing 401331, P.R. China\\
}%

\author{Jin Li}%
\email{cqujinli1983@cqu.edu.cn}
\affiliation{%
    Department of Physics,Chongqing University, 
Chongqing 401331, P.R. China\\
}%
\affiliation{%
Chongqing Key Laboratory for Strongly Coupled Physics, Chongqing University, Chongqing 401331, P.R. China\\
}%
\affiliation{%
Institute of Advanced Interdisciplinary Studies,Chongqing University, Chongqing 401331, China\\
}%

\author{Brendan Mccane}%
\affiliation{%
    School of Computing, University of Otago, Otago 9016, New Zealand\\
}%

\author{Nan Yang}%
\affiliation{%
    Department of Electronical Information Science and Technology, Xingtai University, Xingtai 054001, P.R. China\\
}%
\affiliation{%
Chongqing Key Laboratory for Strongly Coupled Physics, Chongqing University, Chongqing 401331, P.R. China\\
}%

\author{Xianghe Ma}%
\affiliation{%
    Department of Physics,Chongqing University, 
Chongqing 401331, P.R. China\\
}%
\affiliation{%
Chongqing Key Laboratory for Strongly Coupled Physics, Chongqing University, Chongqing 401331, P.R. China\\
}%

\author{Borui Wang}%
\affiliation{%
    Department of Earth and Sciences, Southern University of Science and Technology, 
Shenzhen 518055, P.R. China\\
}%

\author{Minghui Zhang}%
\affiliation{%
    Department of Physics, Southern University of Science and Technology, 
Shenzhen 518055, P.R. China\\
}%

\date{\today}

\begin{abstract}

Gravitational waves from binary neutron star mergers provide insights into dense matter physics and strong-field gravity, but waveform modeling remains computationally challenging.  
We develop a deep generative model for gravitational waveforms from binary neutron star mergers, covering the late inspiral, merger, and ringdown, incorporating  precession and tidal effects. Using the conditional autoencoder, our model efficiently generates waveforms with high precision across a broad parameter space, including component masses $(m_1, m_2)$, spin components $(S_{1x}, S_{1y}, S_{1z}, S_{2x}, S_{2y}, S_{2z})$ and tidal deformability $(\Lambda_1, \Lambda_2)$. Trained on $1\times10^6$ waveforms from the IMRPhenomXP\_NRTidalv2 waveform model, our model achieves a mean mismatch of $2.13\times10^{-3}$.  
The model accelerates waveform generation. For a single sample, it requires  $0.12$ seconds (s), compared to $0.66$ s for IMRPhenomXP\_NRTidalv2 making it approximately 5 times faster. When generating 1000 waveforms, the network completes the task in $0.75$ s, while IMRPhenomXP\_NRTidalv2 requires $7.12$ s, making it approximately 10 times faster.
This speed advantage enables rapid parameter estimation and real-time gravitational wave searches. With higher precision, it will support low-latency detection and broader applications in multi-messenger astrophysics.

\end{abstract}

\maketitle

\section{Introduction}

Since the first direct detection of the binary black hole (BBH) merger GW150914 by LIGO and Virgo in 2015 \cite{abbott2016observation}, gravitational wave (GW) astronomy has entered a new era, enabling direct exploration of extreme astrophysical phenomena. With continuous advancements in detector sensitivity, an increasing number of BBH and binary neutron star (BNS) mergers have been observed \cite{abbott2017gw170817,abbott2020gw190425,abbott2021observation}, providing valuable constraints on the equation of state (EoS) of nuclear matter and insights into tidal interactions in neutron stars \cite{abbott2017gw170817, radice2018binary}. BNS mergers serve as natural laboratories for testing general relativity in the strong-field regime and probing high-density nuclear matter \cite{sagunski2018neutron}. In particular, measurements of tidal deformability impose stringent constraints on the nuclear EoS, shedding light on neutron star structure and ultra-dense matter properties \cite{flanagan2008constraining, hinderer2008tidal}. Additionally, multi-messenger observations, which combine gravitational waves with electromagnetic counterparts, offer an independent method for measuring cosmological parameters, including the Hubble constant \cite{ligo2017gravitational, hotokezaka2018neutron}.  

Due to the significance of BNS systems, accurately modeling gravitational waveforms from their mergers is essential for both detection sensitivity and parameter estimation precision. GW searches rely on matched filtering techniques, which require highly accurate waveform templates, while extracting key physical parameters—such as masses, spins, and tidal deformabilities—demands waveform models with high precision. However, modeling BNS waveforms remains challenging due to complex physical effects, particularly spin precession and tidal interactions~\cite{bernuzzi2012tidal,hannam2014simple,bohe2017improved}. Waveform modeling has progressed from computationally expensive numerical relativity (NR) simulations~\cite{centrella2010black,baumgarte2003numerical,loffler2012einstein}, which solve Einstein’s equations directly, to more efficient semi-analytical methods. NR simulations yield high-precision waveforms by capturing strong-field, nonlinear effects but are too costly for large-scale parameter-space studies. Post-Newtonian (PN) approximations~\cite{blanchet2014gravitational,einstein1938gravitational} describe the inspiral phase under weak-field, slow-motion assumptions but lose accuracy near merger. The effective-one-body (EOB) approach~\cite{buonanno1999effective,damour2008faithful,buonanno2000transition} improves upon PN by mapping the two-body problem to a single-body motion in a modified spacetime, and with NR calibration, balances accuracy and efficiency. Phenomenological models (IMRPhenom)~\cite{hannam2014simple,khan2019phenomenological,ajith2007phenomenological,khan2016frequency} further enhance efficiency by fitting frequency-domain templates to extensive datasets, enabling rapid waveform generation and facilitating large-scale searches. Despite advancements, challenges remain in computational cost, accuracy, and full parameter-space coverage~\cite{Doctor2017Statistical,Field2013Fast,Nagar2018Efficient}, limiting real-time GW detection and precise parameter estimation.

The rapid development of deep learning has introduced an efficient and accurate approach to gravitational waveform modeling \cite{grimbergen2024generating,freitas2024nrsurnn3dq4,liao2021deep,shi2024rapid}. With strong nonlinear fitting capabilities and high computational efficiency \cite{He2021Principle}, deep learning enables high-precision waveform generation at significantly reduced cost. George et al. \cite{george2018deep} first applied deep learning to BBH waveforms and achieved real-time performance beyond traditional methods. Schmidt et al. \cite{schmidt2021machine} used PCA with machine learning to reduce the dimensionality of EOB waveforms and improve efficiency. Dax et al. \cite{dax2025real} accelerated waveform generation using the JAX framework for highly efficient computation and real-time inference.
Beyond BBH systems, deep learning has been applied to BNS and EMRI waveform modeling. Whittaker et al. \cite{whittaker2022machine} used a cVAE to model post-merger signals with EoS uncertainties. Chua et al. \cite{chua2020rapid} combined reduced-order modeling with deep learning to accelerate EMRI waveform generation, reducing the cost by over four orders of magnitude. These works show that deep learning accelerates waveform generation and handles high-dimensional parameter spaces effectively.

Despite progress in deep learning-based waveform modeling, most existing models focus on BBH systems or simplified BNS mergers, with precession and tidal effects remaining underexplored. To address this, we propose a Conditional Autoencoder (cAE) model for rapid BNS waveform generation, with applications in GW data analysis.
Our model efficiently generates waveforms conditioned on system parameters $(\Theta)$, including component masses $(m_1, m_2)$, spin components $(S_{1x}, S_{1y}, S_{1z}, S_{2x}, S_{2y}, S_{2z})$, and tidal deformability $(\Lambda_1, \Lambda_2)$, while capturing the high-dimensional evolution of GW signals. Trained on a dataset of $3\times10^5$ BNS waveforms from the IMRPhenomXP\_NRTidalv2 \cite{dietrich2019improving} model, it incorporates both precession and tidal effects.
To enhance learning efficiency, we adopt the amplitude ($A$)-phase ($\Phi$) representation, where $h_+(t)$ and $h_\times(t)$ are expressed in terms of amplitude and phase independently, to reduce data oscillation. The cAE architecture employs a dual-encoder structure, separately encoding physical parameters and waveform data, which are mapped in latent space before reconstruction. By relying solely on forward propagation, cAE achieves high acceleration in large-scale waveform generation.
Benchmark tests show that generating a single waveform takes 0.12 s with cAE on an NVIDIA A800 80 GB GPU, compared to 0.66 s with IMRPhenomXP\_NRTidalv2 on two Intel Xeon Silver 4214R CPUs (24 cores), corresponding to a speed-up of about 5 times; for a batch of 10³ waveforms, cAE needs 0.75 s compared with 7.12 s, yielding roughly 10 times speed-up.
The model’s accuracy is evaluated through waveform overlap calculations, yielding an average mismatch $2.13\times10^{-3}$, corresponding to accuracy 99.79\%. These results demonstrate that the proposed model enables efficient, accurate, and scalable BNS waveform generation with precession and tidal effects, making it well-suited for real-time signal detection and parameter estimation.

The structure of the article is as follows: Section~\ref{sec:Data simulation} describes the waveform representation and the construction of our dataset. Section~\ref{sec:Autoencoder And Conditional Autoencoder} introduces the fundamental concepts of autoencoders and presents the architecture and hyperparameter settings of our neural network. Section~\ref{sec:Training and Verification} details the model training and validation process. Section~\ref{sec:Results and Analysis} evaluates the accuracy and generation efficiency of our model. Finally, Section~\ref{sec:Summary and Discussion} provides a summary and discusses future research directions.

\section{Data simulation}
\label{sec:Data simulation}
This study constructs a dataset of simulated BNS gravitational waveforms to train a cAE. The dataset spans a broad range of physical parameters ($\Theta$), including component masses, spins, and tidal deformability, and provides the corresponding amplitude and phase representations. This formulation enhances the efficiency of deep learning models in capturing waveform structures and their dependencies on $\Theta$.

\subsection{Waveform Representation}

Gravitational waves are typically characterized by two polarization components, $h_+$ and $h_\times$, expressed as  
\begin{equation}
    h(t) = h_{+}(t) + i h_{\times}(t).
\end{equation}
However, directly learning $h_+(t)$ and $h_\times(t)$ in the time domain is computationally demanding and may hinder training convergence due to waveform complexity. To improve learning efficiency, we adopt an amplitude-phase representation, where $h_+$ is treated as the real part and $h_\times$ as the imaginary part. The corresponding amplitude $A(t)$ and cumulative phase $\Phi(t)$ are given by  
\begin{equation}
    A(t) = \sqrt{h_+^2 + h_\times^2}, \quad
    \Phi(t) = \tan^{-1} \left(\frac{h_\times}{h_+} \right).
    \label{eq:amp_phase_decomposition}
\end{equation}
This representation reduces data Oscillatory while enhancing physical interpretability. The amplitude $A(t)$ captures the overall intensity variation of the gravitational wave, while the cumulative phase $\Phi(t)$ describes its temporal evolution, offering a depiction of the underlying dynamics.  
To further standardize waveform properties, we apply phase normalization, 
\begin{equation}
    \Phi(t) = \Phi(t) - \Phi(t_0),
\end{equation}
which aligns all waveforms to an initial phase of zero. This adjustment improves dataset consistency and stabilizes model training by minimizing phase discrepancies across waveforms, facilitating a more effective learning of parameter dependencies in waveform evolution.

\subsection{System Parameter Selection}

Previous studies on gravitational waveform modeling have primarily focused on BBH systems, while investigations of BNS  waveforms remain relatively limited. Most existing deep learning models assume binary neutron stars with spins aligned to the orbital angular momentum, and thus do not systematically account for spin precession effects. Additionally, although tidal deformation can significantly affect the phase evolution of BNS waveforms, it is often simplified using point-mass approximations, which may result in the omission of tidal contributions.
In this work, we fix the luminosity distance to 1 Mpc, and set both the inclination angle and the coalescence phase to zero. The remaining  parameters are listed in TABLE~\ref{table:parameters}, where the spin magnitude ranges are chosen based on the benchmark settings provided in~\cite{colleoni2025new}.

\begin{table*}[ht]
  \centering
  \caption{Range of the sampling parameters $\Theta$ for the BNS training set (with $m_{2}<m_{1}$).}
  \label{table:parameters}
  \setlength{\tabcolsep}{30pt} 
  \begin{tabular}{ccc}
    \toprule
    $\Theta$ & Description                       & Range                  \\
    \midrule
    $m_{1}$            & Primary mass                              & Uniform$[1,\,3]\;M_{\odot}$       \\
    $m_{2}$            & Secondary mass                            & Uniform$[1,\,3]\;M_{\odot}$       \\
    $|\vec{s}_1|,\;|\vec{s}_2|$ & Spin magnitudes of two neutron stars         & Uniform$[0,\,0.5]$               \\
    $\vec{s}_{1} /\left|\vec{s}_{1}\right|,\vec{s}_{2} /\left|\vec{s}_{2}\right|$     & Spin directions (unit vectors)              & Isotropic over 3D sphere         \\
    $\Lambda_{1},\;\Lambda_{2}$ & Tidal deformabilities of two neutron stars  & Uniform$[0,\,500]$               \\
    \bottomrule
  \end{tabular}
\end{table*}

\subsection{Construction of Dataset}
\label{Construction of Dataset}

We construct a dataset of gravitational waveforms for BNS systems to train and evaluate our conditional autoencoder model. The training set consists of $1\times10^{6}$ samples, while the test set comprises $1\times10^{5}$ samples. Both sets are generated using the IMRPhenomXP\_NRTidalv2 waveform model~\cite{colleoni2025new}, which incorporates precession and tidal effects.

The data generation process includes parameter sampling, waveform computation, preprocessing, and normalization. The sampled parameters $\Theta$ include component masses, spin vectors, and tidal deformabilities, with ranges summarized in TABLE~\ref{table:parameters}. Component masses $m_1$ and $m_2$ ($m_2 < m_1$) are independently drawn from a uniform distribution over $[1,\,3]\,M_{\odot}$, with values reordered post-sampling to ensure $m_1 > m_2$. The dimensionless spin magnitudes $|\vec{s}_1|$ and $|\vec{s}_2|$ are sampled uniformly in $[0,\,0.5]$, and their directions are drawn from an isotropic distribution on the unit sphere to allow for generic spin precession. Tidal deformability parameters $\Lambda_1$ and $\Lambda_2$ are sampled uniformly in $[0,\,500]$.

Both training and test sets are constructed using the same stochastic (non-grid) sampling strategy without predefined step sizes. This randomized approach avoids artifacts introduced by grid-based sampling, enhances waveform diversity, and promotes broad coverage of the high-dimensional parameter space—factors essential for improving generalization and avoiding overfitting in deep learning models. The distributions of sampled parameters in both the training and testing sets are visualized in Appendix~\ref{sec:appendix}, FIGs.~\ref{fig:training_parameter_distribution} and \ref{fig:testing_parameter_distribution}, respectively.

The time-domain GW signals $h_+(t)$ and $h_\times(t)$ are computed using \texttt{pycbc.waveform.get\_td\_waveform}~\cite{pycbc}, followed by trimming to remove leading and trailing zero values. Each waveform is standardized to a fixed duration of $2$ seconds with a sampling rate of 4096 Hz. This segment is taken from the final two seconds before the end of the ringdown, ensuring that the dataset captures the complete merger and ringdown stages while also covering part of the inspiral. The chosen duration is sufficiently long to encompass the stage of prominent tidal effects preceding the merger~\cite{pal2025tidal}, enabling the model to learn the characteristic evolution of waveforms across different dynamical regimes.

\begin{figure*}[!ht]
    \centering
    \includegraphics[width=1\textwidth]{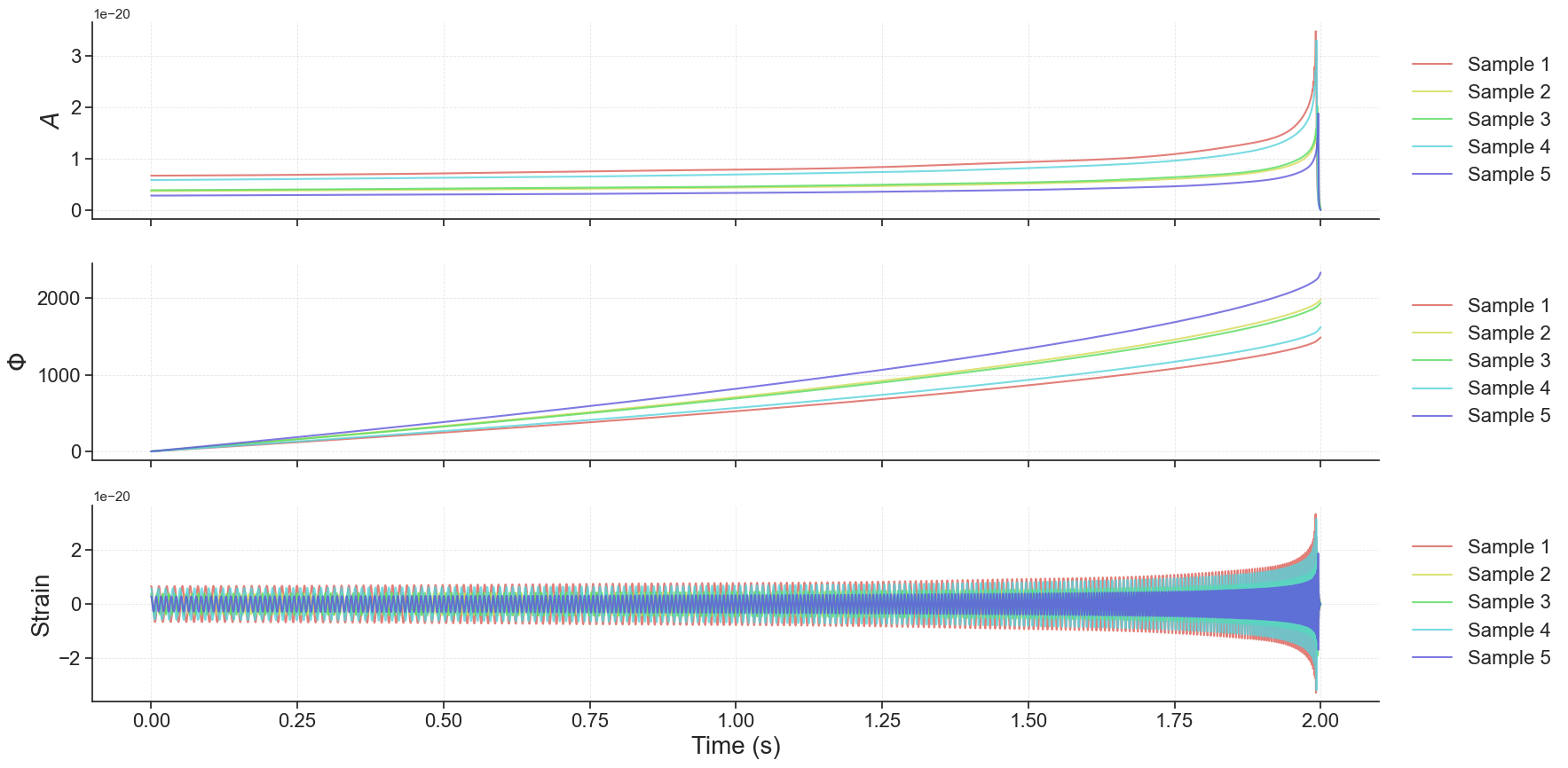}
    \caption{Input samples in time domain. Top: amplitude curve $A(t)$; middle: phase curve $\Phi(t)$; bottom: waveform strain $h(t)$.}
    \label{fig:samples}
\end{figure*}

To maintain data consistency, waveforms are to a fixed duration of $2$ seconds with a sampling rate of 4096 Hz. Each waveform segment is taken from the 2 seconds before the end of the ringdown, ensuring that the dataset captures the complete merger and ringdown stages while also covering part of the inspiral. The 2-second simulated waveform is sufficiently long to encompass the stage of tidal effects preceding the merger \cite{pal2025tidal}.

The dataset consists of three components: $X_{\text{train}}$, which contains the BNS system's $\Theta$ as conditional inputs; $y_{A}$, representing the amplitude data; and $y_{\Phi}$, representing the phase data.  
During cAE training, we normalize the data to ensure stability and consistency. The $X_{\text{train}}$ are processed using Min-Max Normalization \cite{rahmad2024comparative}, which rescales the data to the range $[0,1]$, ensuring a uniform scale across different parameters and improving training stability:  
\begin{equation}
	X' = \frac{X - X_{\min}}{X_{\max} - X_{\min}},
\end{equation}
where $X$ represents the original data, $X_{\min}$ and $X_{\max}$ are the minimum and maximum values of each parameter, and $X'$ is the normalized data.  
For $y_{A}$ and $y_{\Phi}$, we apply standardization \cite{singh2021exploring}, which ensures zero mean and unit variance to eliminate scale differences and enhance training stability:  
\begin{equation}
	X' = \frac{X - \mu}{\sigma},
\end{equation}
where $\mu$ and $\sigma$ denote the mean and standard deviation of the data, respectively.  
Since the amplitude and phase exhibit distinct evolution patterns, two separate cAEs are trained: one for learning a compact representation and reconstruction of $y_{A}$, and another for modeling $y_{\Phi}$. During training, $X_{\text{train}}$ is provided as a conditional variable to the autoencoder, ensuring that the latent representation $z$ effectively captures the dependence of waveforms on  $\Theta$. 

\section{Autoencoder And Conditional Autoencoder}
\label{sec:Autoencoder And Conditional Autoencoder}
The autoencoder (AE) is an unsupervised learning model widely used for data dimensionality reduction, feature extraction, and generative tasks. It consists of an encoder and a decoder, learning a low-dimensional representation of the data by minimizing reconstruction error. The variational autoencoder (VAE) extends this model by introducing probabilistic modeling, enforcing a smoother latent variable distribution, which enhances the generative capability. The cAE further incorporates external conditional constraints, enabling the model to generate samples corresponding to specific data distributions based on input conditions, making it particularly relevant for GW waveform modeling.  

This section first introduces the fundamental concepts of AE, VAE, and cAE, discussing their applicability to GW waveform generation. Subsequently, we provide a detailed description of the proposed cAE-based waveform generation model, including the separate cAE architectures designed for phase and amplitude modeling, along with their respective hyperparameter settings.

\subsection{Concepts of Autoencoder and Conditional Autoencoder}

\begin{figure*}[!ht]
    \centering
    \begin{subfigure}[b]{0.3\textwidth}
        \centering
        \includegraphics[width=\textwidth]{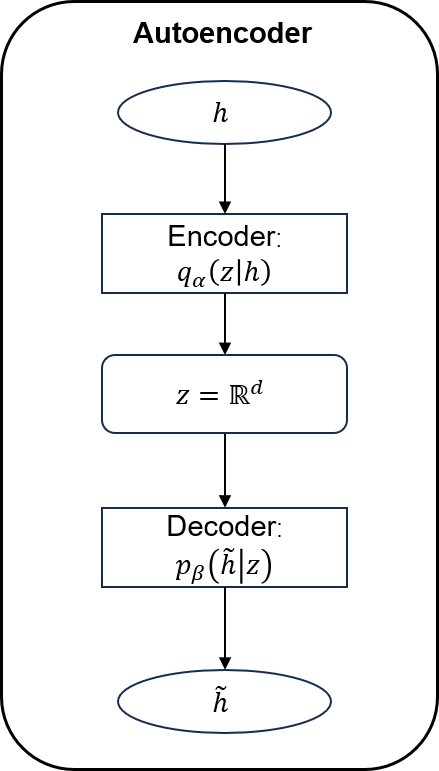}
        \caption{}
        \label{fig:ae}
    \end{subfigure}
    \hspace{0.05\textwidth}
    \begin{subfigure}[b]{0.3\textwidth}
        \centering
        \includegraphics[width=\textwidth]{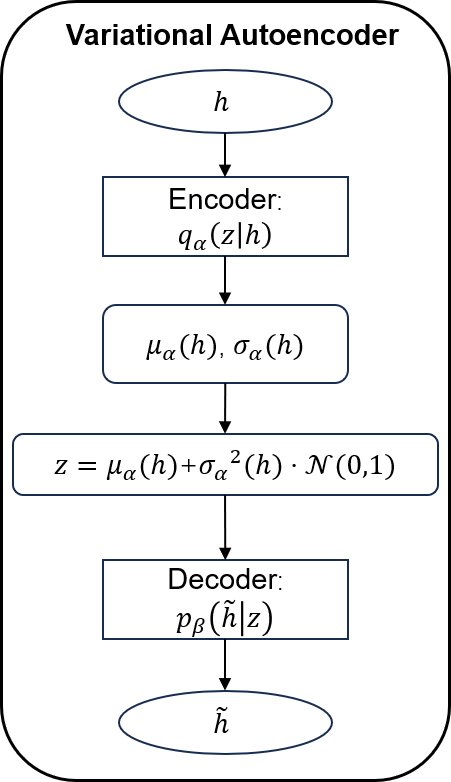}
        \caption{}
        \label{fig:vae}
    \end{subfigure}
    \caption{(a) Structure of an AE. (b) Structure of a variational VAE.}
    \label{fig:ae_vae_combined}
\end{figure*}

We employ autoencoders (AEs) \cite{hinton1993autoencoders} to reduce the dimensionality of complex GW waveforms (Amplitude $A$ and Phase $\Phi$) while ensuring accurate reconstruction. As shown in FIG.~\ref{fig:ae}, an AE consists of an input  $h^{(i)}$ ($A$ or $\Phi$), an encoder $q_{\alpha}(z \mid h)$, a latent variable $z^{(i)} \in \mathbf{R}^{d}$, and a decoder $p_{\beta}({\hat{h}} \mid z)$. The encoder projects the input ($A$ or $\Phi$)  into a lower-dimensional latent space $z^{(i)}$, where the dimension $d$ of $z^{(i)}$ can be adjusted based on specific task requirements. The decoder then reconstructs the  $\hat{h}^{(i)}$ from the latent representation through an upsampling process. The $\alpha$  and $\beta$ refers to the learned model parameters, such as weights and biases, obtained after training. To measure the similarity between the reconstructed  $\hat{h}^{(i)}$ and the target  $h^{(i)}$, the AE employs the mean squared error (MSE) as the reconstruction loss:
\begin{equation}
	L_{\text{MSE}} = \frac{1}{N} \sum_{i=1}^{N} \left\| h^{(i)} - \hat{h}^{(i)} \right\|^2,
\end{equation}
where $N$ is the total number of training samples, $h^{(i)}$ represents the target ($A$ or $\Phi$), and $\hat{h}^{(i)}$ is the reconstructed $A$ or $\Phi$. By minimizing $L_{\text{MSE}}$, the model updates its weights and biases, ensuring that $\hat{h}^{(i)}$ closely approximates $h^{(i)}$.  Unlike traditional linear methods such as PCA \cite{mackiewicz1993principal}, autoencoders (AEs) can capture the nonlinear features of GW signals more effectively \cite{cacciarelli2023hidden}. While PCA is efficient for simple signals, its linear projections may miss important features in the nonlinear phases of GW evolution \cite{bloomer2010principal,liborio2022principal}. In contrast, AEs reduce dimensionality through nonlinear mappings, preserving key physical features such as orbital dynamics, tidal effects, and ringdown. This enables better generalization, parameter recovery, and interpolation across the waveform space \cite{ladjal2019pca,nousi2022autoencoder}.
FIG.~\ref{fig:vae} illustrates the structure of a VAE \cite{kingma2019introduction}. Unlike standard AEs, VAEs introduce probabilistic modeling between the encoder and decoder, ensuring that the latent variable $z^{(i)}$ is not a fixed deterministic value but is instead sampled from a distribution defined by the encoder’s output mean $\mu_\alpha(h^{(i)})$ and variance $\sigma_\alpha^2(h^{(i)})$. Specifically, VAEs utilize the reparameterization trick to obtain latent variables:
\begin{equation}
	z^{(i)} = \mu_{\alpha}(h^{(i)}) + \sigma_{\alpha}(h^{(i)}) \cdot \epsilon, \quad \epsilon \sim \mathcal{N}(0, I).
\end{equation}
This approach allows gradients to propagate through the sampling process, making it possible to optimize the network using gradient-based methods. The VAE training objective consists of the reconstruction loss and the Kullback-Leibler (KL) divergence loss \cite{asperti2020balancing,prokhorov2019importance}. The reconstruction loss measures the difference between the decoder's output $\hat{h}^{(i)}$ and the input waveform $h^{(i)}$, typically computed using the negative log-likelihood:
\begin{equation}
	L_{\text{recon}} = \mathbb{E}_{q_{\alpha}(z \mid h)} \left[ -\log p_{\beta}(\hat{h} \mid z) \right].
\end{equation}
The KL divergence loss ensures that the learned latent variable distribution $q_{\alpha}(z \mid h)$ approximates a predefined prior distribution, typically a standard normal distribution $p(z) = \mathcal{N}(0, I)$ where \( I \) is the identity matrix:
\begin{equation}
	L_{\text{KL}} = D_{\mathrm{KL}} \left( q_{\alpha}(z \mid h) \| p(z) \right).
\end{equation}
The final VAE objective function is given by:
\begin{equation}
	L_{\text{VAE}} = L_{\text{recon}} + \kappa L_{\text{KL}},
\end{equation}
where the hyperparameter $\kappa$ controls the weight of the KL divergence loss, regulating the structure of the latent space.

Although traditional AEs and VAEs perform well in capturing the low-dimensional structure and nonlinear features of data, their generative processes typically rely solely on the data itself. As a result, they lack the capacity to explicitly incorporate known physical priors into the latent representations. In other words, standard AE/VAE models in unsupervised learning tend to capture the dominant variations in the data, but they cannot guarantee that the generated waveforms strictly adhere to physical constraints. 
To address this limitation and further enhance the physical interpretability and controllability of waveform generation, we use the cAE~\cite{zhang2021conditional}. In the cAE model, additional physical parameters $\Theta$ (as shown in TABLE \ref{table:parameters}) are incorporated as conditional inputs and jointly mapped with waveform data into a low-dimensional latent space. In this manner, the cAE not only inherits the advantages of AE/VAE in nonlinear dimensionality reduction and data reconstruction, but also enables the explicit embedding of physical constraints into the latent variables, thereby generating waveforms that better reflect realistic astrophysical properties.

In the following, we provide a detailed description of our cAE model architecture. 

\subsection{Architecture and Hyperparameters of the Model}

\begin{figure*}[htbp]
    \centering
    \includegraphics[width=0.95\textwidth]{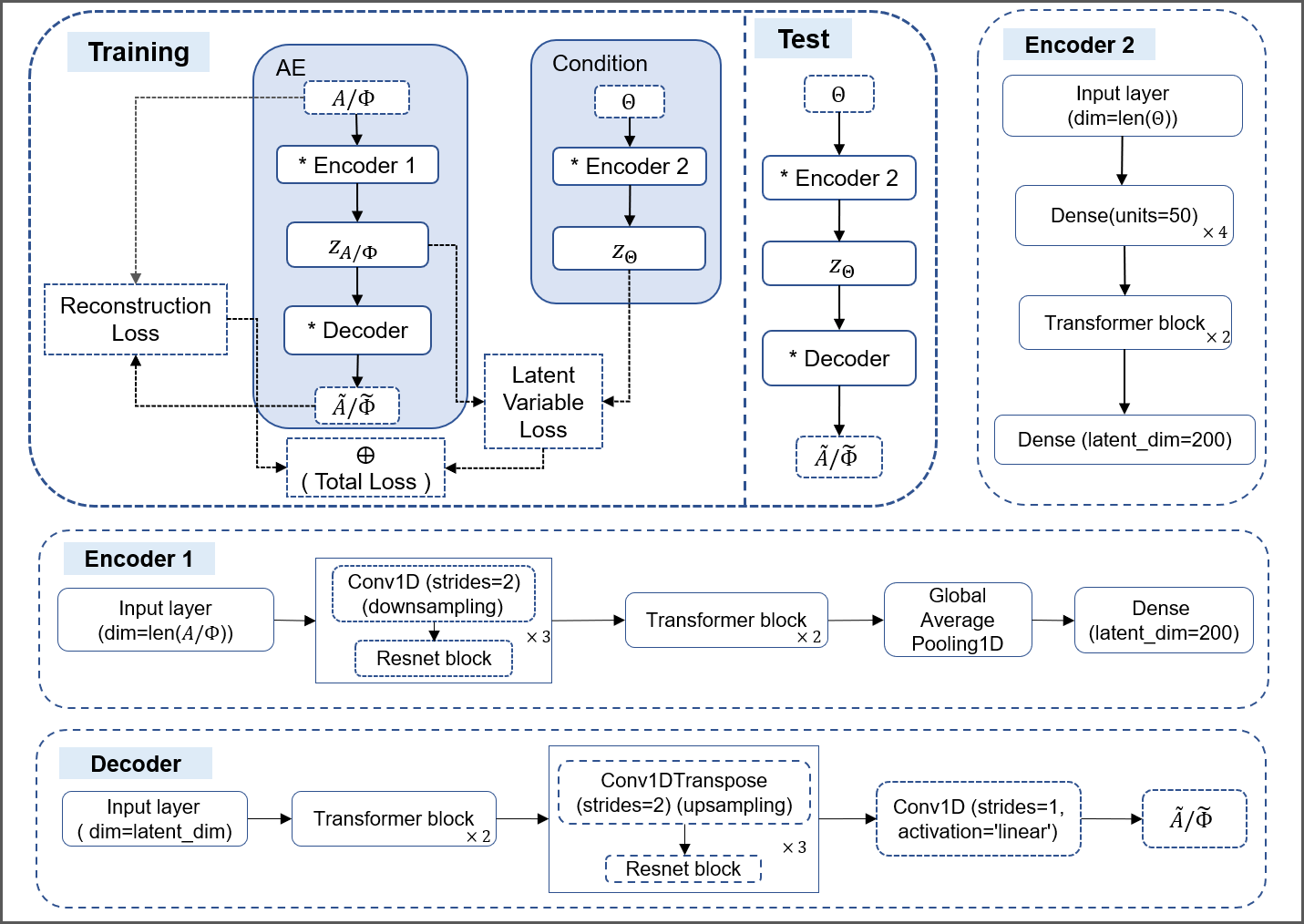}
    \caption{Training and testing architecture of our model, along with detailed structures of individual components, the structure of the Transformer and ResNet block are shown in Appendix~\ref{sec:appendix} (FIG~\ref{fig:resnet_transformer}).}
    \label{fig:total_network}
\end{figure*}

Our study employs the cAEs to model the amplitude and phase of GW waveforms from BNS mergers. The model consists of two independent cAEs, each responsible for learning a low-dimensional representation of either the amplitude $y_{A}$ or phase $y_{\Phi}$ and reconstructing waveforms conditioned on  $\Theta$. In our implementation, the dimensionality of the latent representation is set to 300. Each cAE comprises two encoders (Encoder 1 and Encoder 2) and a decoder. Encoder 1 processes the amplitude or phase data, while Encoder 2 encodes the $\Theta$, and the two latent representations are combined in the latent space before being mapped back to a complete waveform by the decoder. The overall model architecture is shown in FIG. \ref{fig:total_network}, where Training and Test sections correspond to the training and inference workflows, while the remaining sections detail the structural components.

As shown in FIG.~\ref{fig:total_network}, our model consists of two main branches: one for waveform encoding and reconstruction, and one for encoding $\Theta$. The goal is to align latent representations from both branches while ensuring accurate waveform reconstruction.

\textbf{Encoder 1} is used independently for amplitude and phase inputs. Each encoder processes a normalized 1D data through a Conv1D layer with stride 2, followed by three ResNet blocks for local feature extraction and two Transformer blocks for capturing long-range dependencies. A global average pooling layer compresses the temporal dimension, and a final dense layer maps the features into a  latent space ($z_{A}$ or $z_{\Phi}$).

\textbf{Encoder 2} takes $\Theta$ (TABLE~\ref{table:parameters}) as input. These are passed through four fully connected layers, followed by two Transformer blocks. The output is projected into the same latent space as Encoder 1, producing $z_{\Theta}$.

\textbf{Decoder} takes the latent variable from Encoder 1 and reconstructs the waveform. It  expands the latent dimension via a dense layer, then applies two Transformer blocks and three upsampling ResNet blocks using transposed Conv1D layers, recovering the waveform shape.

\textbf{Latent variables and loss functions:} During training, both the waveform and the $\Theta$ are encoded into their respective latent representations, denoted as $z_{A/\Phi}$ and $z_{\Theta}$. To ensure that the decoder can accurately reconstruct the input waveform and that both latent spaces are aligned, we define two loss components:
First, the \textit{reconstruction loss} is computed as the Mean Absolute Error (MAE) between the input waveform $x$ and the reconstructed waveform $\hat{h}$:
\begin{equation}
    L_{\text{rec}} = \frac{1}{N} \sum_{i=1}^{N} \left\| h^{(i)} - \hat{h}^{(i)} \right\|_\text{MAE}.
\end{equation}
Second, the \textit{latent consistency loss} penalizes the difference between the latent vector produced by Encoder 1 ($A/\Phi$) and Encoder 2 ($\Theta$):
\begin{equation}
    L_{\text{latent}} = \frac{1}{N} \sum_{i=1}^{N} \left\| z_{{A/\Phi}}^{(i)} - z_{\Theta}^{(i)} \right\|_\text{MSE}^2.
\end{equation}
The total loss is then defined as a weighted sum of the reconstruction and latent consistency losses, as shown in Eq.~\ref{eq:total_loss}:
\begin{equation}
    L_{\text{total}} = L_{\text{rec}} + \lambda L_{\text{latent}},
    \label{eq:total_loss}
\end{equation}
where $\lambda$ is a balancing coefficient. In our study, we set $\lambda = 1$ to equally weight reconstruction accuracy and latent alignment.

With the model architecture defined, we describe the training procedure and validation setup used to evaluate the model's performance in the following section.

\section{Training and Verification}
\label{sec:Training and Verification}

\subsection{Training}

\begin{figure}[htbp]
    \centering
    \includegraphics[width=0.5\textwidth]{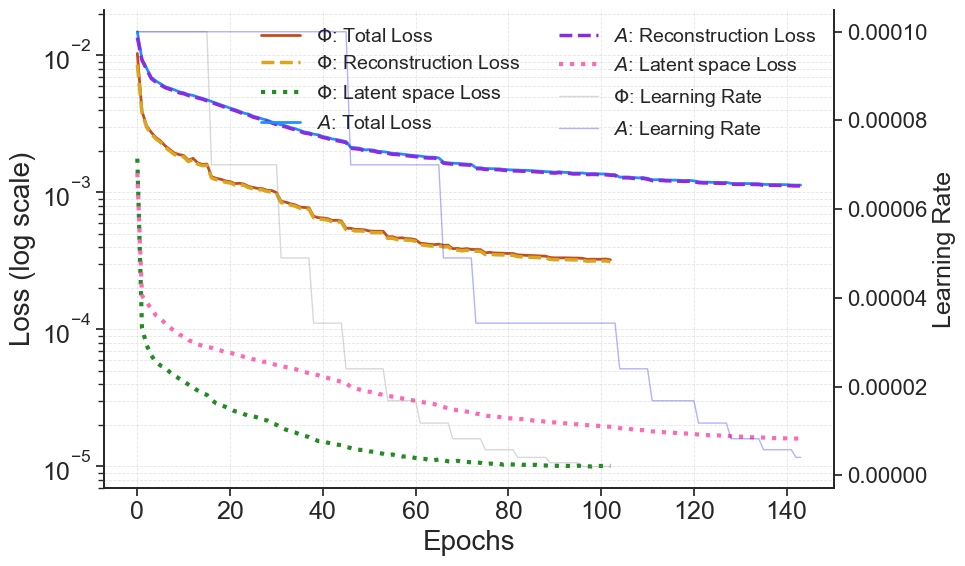}
    \caption{Training and validation loss curves for the amplitude and phase models.}
    \label{fig:training_loss}
\end{figure}

Amplitude and phase models are trained separately on an NVIDIA A800 GPU with 80 GB of memory. The full training process, including both models, takes approximately ten days. The training dataset consists of $1\times 10^6$ waveform samples, split into 90\% for training and 10\% for validation. Before training, input parameters and target waveforms are normalized using Min-Max and Standard scaling, respectively. Data is preprocessed into 10 blocks and fed into the model with a batch size of 128. FIG.~\ref{fig:training_loss} shows the training and validation loss curves for both amplitude and phase models.

Our model is trained using the Adam optimizer with an initial learning rate of $10^{-4}$. To enhance convergence and prevent stagnation, a learning rate scheduling strategy is employed: if the validation loss does not improve for 7 consecutive epochs, the learning rate is reduced by a factor of 0.7, with a floor value of $10^{-8}$. In addition, early stopping is activated when the validation loss shows no improvement over 15 consecutive epochs, and the model is restored to the state with the best validation performance.

\subsection{Verification}
\label{sec:verification}
To evaluate the precision of cAE-generated waveforms, we compute the mismatch between the model-generated waveforms and the IMRPhenomXP\_NRTidalv2 waveforms. The evaluation is based on two metrics: Overlap \cite{owen1996search,owen1999matched} and Mismatch, which quantify the waveform reconstruction quality. The analysis is conducted in the frequency domain after applying a Fourier transform to the time-domain waveforms, allowing for a more effective comparison of waveform similarities. In GW data analysis, the inner product of two waveforms is typically defined as a noise-weighted integral over the frequency domain, incorporating the power spectral density $S_n(f)$:
\begin{equation}
\langle h_1 \mid h_2 \rangle
= 4\text{Re}\int_{f_{\min}}^{f_{\max}}
\frac{\tilde{h}_1(f)\,\tilde{h}_2(f)}{S_n(f)} \,df,
\end{equation}
where $\tilde{h}_1(f)$ and $\tilde{h}_2(f)$ are the Fourier transforms of $h_1(t)$ and $h_2(t)$, respectively, and $S_n(f)$ represents the power spectral density (PSD) of the detector noise. This weighted inner product provides a measure of how well two waveforms match in the presence of detector noise.  We set $S_n(f)$ to the LIGO O4 sensitivity curve \cite{LIGO_T2000012}.

To eliminate the influence of normalization, each waveform is rescaled to satisfy the unit-norm condition:
\begin{equation}
\hat{h}(t) = \frac{h(t)}{\sqrt{\langle h \mid h \rangle}}.
\end{equation}
The Overlap between two waveforms is then computed by maximizing the inner product over different time shifts $t_c$ and phase shifts $\phi_c$:
\begin{equation}
O(h_1, h_2)
= \max_{t_c,\,\phi_c}\,
\frac{\langle \hat{h}_1 \mid \hat{h}_2 \rangle}
{\sqrt{\langle \hat{h}_1 \mid \hat{h}_1 \rangle\,\langle \hat{h}_2 \mid \hat{h}_2 \rangle}}.
\end{equation}
Based on this Overlap metric, the Mismatch is defined as:
\begin{equation}
M(h_1,h_2) = 1 - O(h_1,h_2).
\end{equation}
Here, $M(h_1,h_2)$ quantifies the dissimilarity between the two waveforms, where lower values indicate a higher similarity between the model-generated waveforms and the target physical waveforms.

\section{Results and Analysis}
\label{sec:Results and Analysis}

In this section, we not only compare our proposed conditional autoencoder with traditional waveform approximants, but also with a recent deep learning architecture~\cite{shi2025rapid}, which employs a residual-stacked Multilayer Perceptron (MLP) and Convolutional Neural Network (CNN). Traditional waveform approximants are included as baseline references in both the mismatch analysis and the generation time comparison. We present the generation precision and generation speed of both the cAE and the residual-stacked MLP-CNN model, using the traditional models as a benchmark. In addition, we further analyze the latent space to examine whether physically meaningful correlations are preserved when varying parameters. Moreover, we perform a tidal phasing consistency test by varying tidal deformabilities $(\Lambda_1, \Lambda_2)$ with fixed masses and spins, and compare the cumulative phase evolution against the IMRPhenomXP NRTidalv2 model.

\subsection{Mismatch Evaluation}  
\label{sec:Mismatch Evaluation}
\begin{figure}[htbp]
    \centering
    \includegraphics[width=0.45\textwidth]{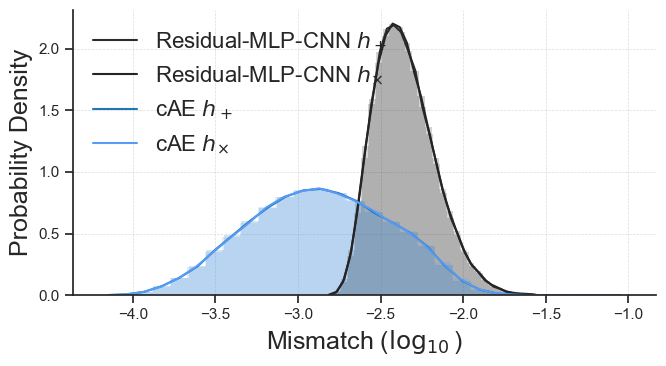}
    \caption{Mismatch distributions of $h_+$ and $h_\times$ for both the cAE and Residual-MLP-CNN models over test set.}
    \label{fig:mismatch}
\end{figure}

To evaluate the accuracy of waveform reconstruction between neural network models and traditional waveform approximants, we compare our conditional autoencoder model with the Residual-MLP-CNN model, whose architecture is reproduced from~\cite{shi2025rapid} and shown in FIG.~\ref{fig:res_mlp_cnn_network} of Appendix~\ref{sec:appendix}. Both models are trained using the same dataset and identical preprocessing pipelines, following the same procedure used for our cAE model. The training strategy including optimizer, learning rate scheduler, early stopping, and callbacks—is also kept consistent with that of cAE model.
Using the test set introduced in Section~\ref{Construction of Dataset}, we evaluate waveform reconstruction accuracy using the mismatch. FIG.~\ref{fig:mismatch} shows the mismatch distributions of $h_+$ and $h_\times$ for both models. The average mismatch is calculated as $\text{mismatch}_\text{avg} = \frac{1}{N} \sum_{i=1}^{N} \text{mismatch}_i$. The average mismatch of the cAE model is $2.13\times10^{-3}$, while that of the Residual-MLP-CNN model is $5.72\times10^{-3}$, indicating that the cAE achieves higher reconstruction accuracy.

We analyze the relationship between the number of orbital cycles and the mismatch by computing the cycle count for  test set. The time-domain signal $s(t)$ undergoes a Hilbert transform to obtain its analytic representation\cite{xiao2012signal}:
\begin{equation}
s_a(t) = s(t) + i\,\mathcal{H}\{s(t)\},
\end{equation}
where $\mathcal{H}\{\cdot\}$ denotes the Hilbert transform. The instantaneous phase is then extracted from the analytic signal:
\begin{equation}
\phi(t) = \arg\bigl(s_a(t)\bigr).
\end{equation}
To eliminate phase discontinuities, we apply phase unwrapping to obtain a monotonic phase function $\tilde{\phi}(t)$ and compute the total phase difference:
\begin{equation}
\Delta \phi = \tilde{\phi}(T) - \tilde{\phi}(0).
\end{equation}
Finally, the orbital cycle number is given by:
\begin{equation}
\text{Cycles} = \frac{\Delta \phi}{2\pi}.
\label{eq:cal_cycle}
\end{equation}

\begin{figure}[htbp]
    \centering
    \includegraphics[width=0.45\textwidth]{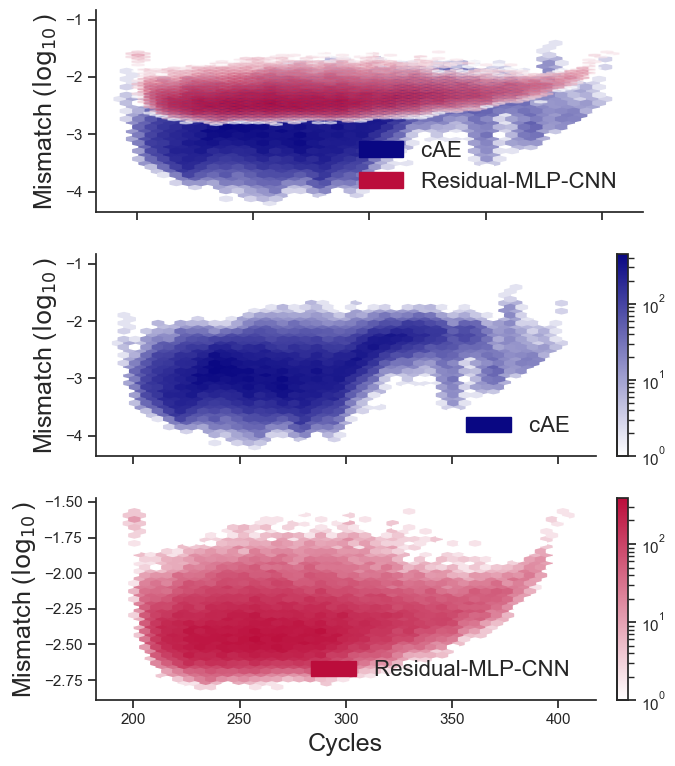}
    \caption{The horizontal axis shows the number of orbital cycles in the test set, and the vertical axis shows the mismatch. Color indicates sample density.}
    \label{fig:cycle_distribution_mismatch}
\end{figure}

FIG.~\ref{fig:cycle_distribution_mismatch} illustrates how the waveform mismatch varies with the number of orbital cycles in the test set. Across a wide range of cycles (200–400), there is no clear trend of mismatch increasing with cycle count. In particular, our cAE exhibits consistently low mismatch values, indicating robust learning performance. The Residual-MLP-CNN model also maintains stable mismatch across the cycle range, though its overall mismatch is higher than that of cAE. These results demonstrate that, given a sufficiently large training dataset ($1\times10^6$ samples), both models are capable of learning accurate waveform representations without degradation at higher cycle counts.
Compared to Residual-MLP-CNN, cAE achieves lower mismatch over nearly the entire cycle range, further highlighting its generalization capacity.

\begin{figure}[htbp]
    \centering
    \includegraphics[width=0.45\textwidth]{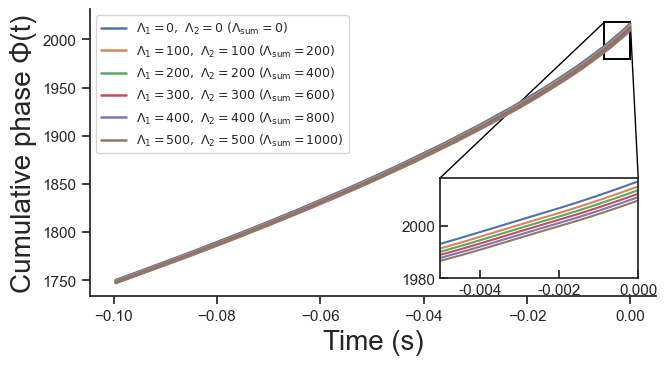}
    \caption{Cumulative phase $\Phi(t)$ evolution for cAE-generated waveforms under different tidal deformabilities $(\Lambda_1,\Lambda_2)$. The inset shows the zoom-in around the merger time.}
    \label{fig:phi_time}
\end{figure}

To further evaluate the physical consistency of our cAE model, we conduct a tidal phasing consistency test. 
In this test, both component masses are fixed at $1.4M_\odot$, and the spins are set to zero for both stars, while the tidal deformabilities $(\Lambda_1, \Lambda_2)$ are varied.
We analyze the cumulative phase evolution $\Phi(t)$ of IMRPhenomXP\_NRTidalv2 waveforms under different tidal deformability parameters. 
Figure~\ref{fig:phi_time} shows the cumulative phase $\Phi(t)$ over time for several representative choices of $(\Lambda_1, \Lambda_2)$. 
As the total tidal deformability $\Lambda_\Sigma = \Lambda_1 + \Lambda_2$ increases, the resulting phase curves exhibit consistent upward shifts, reflecting the acceleration of phase evolution due to stronger tidal effects.

\begin{figure}[htbp]
    \centering
    \includegraphics[width=0.45\textwidth]{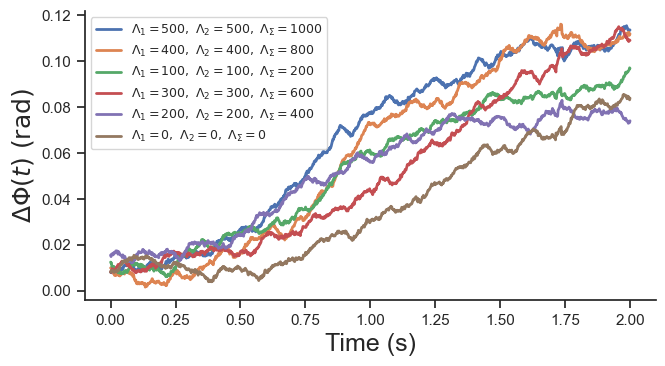}
    \caption{Phase difference $\Delta \Phi(t)$ between the cAE model and IMRPhenomXP\_NRTidalv2 over time for different total tidal deformabilities $\Lambda_\Sigma$.}
    \label{fig:delta_phi_time}
\end{figure}

In FIG.~\ref{fig:delta_phi_time}, we present the phase difference $\Delta \Phi(t)$ between waveforms generated by our cAE model and the IMRPhenomXP\_NRTidalv2. As time increases, the cumulative phase difference gradually grows, reflecting the accumulation of small discrepancies during the inspiral evolution. Nevertheless, the absolute magnitude of $\Delta \Phi(t)$ remains within a relatively small range throughout the entire duration time, indicating that the cAE maintains high accuracy  in phase reconstruction. Importantly, when keeping the component masses and spins fixed and varying only the tidal deformabilities $(\Lambda_1, \Lambda_2)$, we observe a small upward trend in the mismatch as $\Lambda_\Sigma$ increases. This behavior is physically consistent, since stronger tidal effects lead to larger phase shifts in binary neutron star waveforms. However, the mismatch stays at a low level, showing that our cAE can model tidal effects reliably and accurately.

\begin{figure}[htbp]
    \centering
    \includegraphics[width=0.45\textwidth]{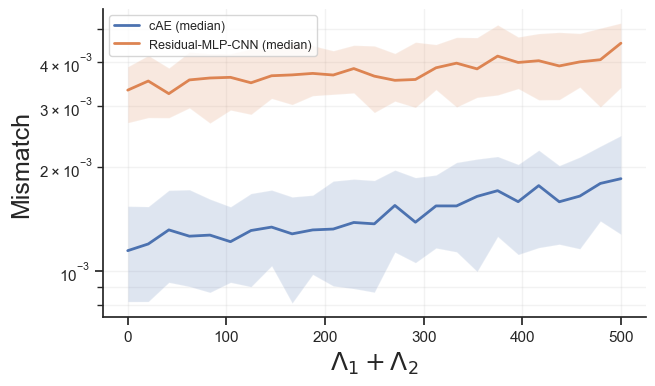}
    \caption{Mismatch between cAE and IMRPhenomXP\_NRTidalv2 waveforms across different values of total tidal deformability $\Lambda_\Sigma = \Lambda_1 + \Lambda_2$. The shaded regions indicate the spread of mismatch values within approximately one standard deviation around the median at each total tidal deformability $\Lambda_\Sigma$.}
    \label{fig:mismatch_total_lambda}\end{figure}

To further verify the robustness of the cAE in the strong tidal regime, we examined how the waveform mismatch changes with the total tidal deformability $\Lambda_\Sigma = \Lambda_1 + \Lambda_2$, and compared it with the Residual-MLP-CNN model.
As shown in FIG.~\ref{fig:mismatch_total_lambda}, both models exhibit a small increasing trend in mismatch as $\Lambda_\Sigma$ grows, which is consistent with the cumulative phase shifts induced by tidal effects.
However, the cAE maintains its mismatch at a consistently lower level (below $2\times 10^{-3}$).

\subsection{Latent Space Analysis}
\label{sec:Latent Space Analysis}

To more systematically examine whether the proposed cAE model captures physically meaningful structures within its latent representation, we have added three controlled experiments to , with the corresponding results shown in FIG. 10. Specifically, in each experiment, we vary only a kind of physical parameter while keeping the others fixed to observe the model's latent vector response: (a) we fix the component masses $m_1 = m_2 = 1.4,M_\odot$ and spins $\chi_{1z} = \chi_{2z} = 0$, and gradually increase the tidal deformability $\Lambda_1 = \Lambda_2$; (b) we fix $\Lambda_1 = \Lambda_2 = 0$ and $\chi_{1z} = \chi_{2z} = 0$, and vary the total mass $M_{\mathrm{tot}}$; (c) we fix the masses and tidal parameters, and vary the total spin $\chi_{\mathrm{tot}}$. For each configuration, we input the physical parameters into the cAE and extract the latent vector $z \in \mathbb{R}^{300}$ from the encoder trained on waveform phase.

To compare how each physical variation affects the latent dimensions, we compute the difference $\Delta z = z_i - z_0$ for each sample relative to the baseline (i.e., the first sample in each set with the smallest parameter value), and sort the latent dimensions by their overall sensitivity. This ensures that dimensions with the most pronounced changes are shown on the left in each figure. Each figure displays 20 samples arranged in a $4\times5$ grid, where each row contains 5 samples corresponding to different physical configurations. For each sample, we obtain a 300-dimensional latent vector from the encoder (trained on waveform phases). The first sample (top-left corner) serves as the baseline, and the remaining 19 samples are compared against it. Specifically, we compute the absolute difference $\Delta z = |z_i - z_0|$ between each sample’s latent vector $z_i$ and the baseline vector $z_0$. Each column represents a latent dimension, and the color intensity reflects the magnitude of $\Delta z$ in that dimension. Brighter colors indicate stronger deviation from the baseline. In addition, a brighter color can be interpreted as a stronger activation of the corresponding latent neuron. A unified color scale is used across all subfigures to ensure consistency in visual interpretation.

\begin{figure*}[htbp]
    \centering

    \begin{subfigure}[b]{0.95\textwidth}
        \centering
        \includegraphics[width=\textwidth]{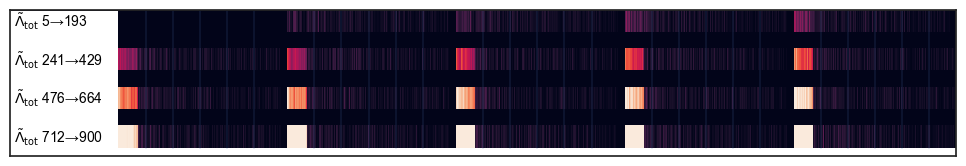}
        \caption{}
        \label{fig:latent_lambda} 
    \end{subfigure}
    
    \begin{subfigure}[b]{0.95\textwidth}
        \centering
        \includegraphics[width=\textwidth]{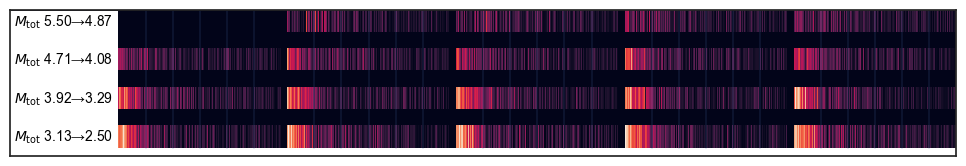}
        \caption{}
        \label{fig:latent_mass}   
    \end{subfigure}
    
    \begin{subfigure}[b]{0.95\textwidth}
        \centering
        \includegraphics[width=\textwidth]{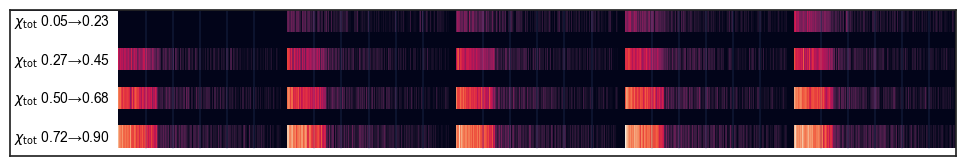}
        \caption{}
        \label{fig:latent_spin}   
    \end{subfigure}

\caption{Latent space analysis of the cAE model under controlled variations of physical parameters. Each subfigure contains 20 samples arranged in a $4\times5$ grid, where each row contains 5 waveform configurations. Each sample is represented by a horizontal strip of 300 cells, corresponding to the 300-dimensional latent vector output by the encoder. Each column thus represents a specific latent dimension across all samples. The color of each cell indicates the relative activation magnitude $\Delta z$ compared to a baseline (the first sample in each subfigure), with brighter colors representing stronger positive deviations. The baseline row appears uniformly dark by construction. All subfigures share a unified color scale for consistent visual comparison. Experimental setups:  
(a) Varying tidal deformabilities ($\Lambda_1=\Lambda_2$) with fixed component masses ($m_1=m_2=1.4M_\odot$) and spins ($\chi_{1z}=\chi_{2z}=0$);  
(b) Varying total mass with fixed spins ($\chi_{1z}=\chi_{2z}=0$) and tidal deformabilities ($\Lambda_1=\Lambda_2=0$);  
(c) Varying total spin $\chi_{\rm tot}$ with fixed component masses and tidal deformabilities.}
\label{fig:latent_space_interpretation}

\end{figure*}

The results show interpretable correlations between physical parameters and latent responses. For example, in the tidal variation experiment, only a small subset of latent dimensions exhibits significant and stable increases in activation, suggesting that the cAE has learned to encode tidal effects—which primarily impact late-inspiral phase evolution—into a compact subspace (see Fig.~\ref{fig:latent_lambda}). In the mass variation experiment, lower total mass leads to both stronger and more widespread activations, consistent with the fact that lower-mass binaries produce longer inspirals with richer phase features (see Fig.~\ref{fig:latent_mass}). In contrast, for spin variation, the response is more diffuse across many latent dimensions, reflecting the more complex and nonlocal influence of spin, which affects both phase evolution and precession modulation(see Fig.~\ref{fig:latent_spin}).

In Appendix~\ref{sec:appendix}, FIGs.~\ref{fig:c_200}-\ref{fig:c_350} illustrate several examples of reconstructed waveforms generated by our conditional autoencoder model for test cases with different physical parameters.

\subsection{Waveform Generation Efficiency}
\label{sec:Waveform Generation Efficiency}

We evaluate waveform generation speed for three categories of models:
(i) our conditional autoencoder,  
(ii) the reproduced Residual-MLP-CNN,
(iii) four CPU-based approximants (SpinTaylorT1, IMRPhenomPv2\_NRTidal, IMRPhenomPv2\_NRTidalv2, and IMRPhenomXP\_NRTidalv2).  
All timings were obtained on an NVIDIA A800 80GB GPU and two Intel Xeon Silver 4214R CPU (24 cores).

\paragraph{Single waveform.}  
    Table~\ref{tab:single_time} shows that our cAE requires 0.1204 s on GPU and 0.2314 s on CPU to generate a single waveform. Compared to traditional CPU-based approximants, the cAE achieves a speed-up of approximately 5–9 times on GPU and remains slightly faster on CPU. When compared to the Residual-MLP-CNN, the two models exhibit nearly identical performance on GPU (0.1204\,s vs. 0.1221 s), and similarly on CPU (0.2314 s vs. 0.1675 s), indicating broadly consistent generation times across both architectures. Overall, both neural models significantly outperform traditional approximants in single waveform generation time.

\begin{table}[htbp]
\centering
\caption{Time to generate a single waveform. 
Entries marked with an asterisk (*) denote AI-based models. 
GPU-based results are obtained on an NVIDIA A800 80GB, while CPU results are from two Intel Xeon Silver 4214R. Traditional approximants are evaluated on CPU only.}
\label{tab:single_time}
\begin{tabular}{l@{\hspace{1cm}}r}
\hline
\textbf{Waveform}                           & \textbf{Time (s)} \\ \hline
*cAE (GPU)                   & 0.1204 \\
*cAE (CPU)                   & 0.2314 \\
*Residual-MLP-CNN (GPU)      & 0.1221 \\
*Residual-MLP-CNN (CPU)      & 0.1675 \\[2pt]
SpinTaylorT1                 & 0.9415 \\
IMRPhenomPv2\_NRTidal        & 1.1026 \\
IMRPhenomPv2\_NRTidalv2      & 0.8383 \\
IMRPhenomXP\_NRTidalv2       & 0.6663 \\
\hline
\end{tabular}
\end{table}

\paragraph{Batched generation.}  
Table~\ref{tab:batch_efficiency} presents waveform generation times for batch sizes ranging from 1 to 1000. The conditional autoencoder (cAE) and Residual-MLP-CNN exhibit similar performance across all batch sizes. On GPU, their generation times are nearly identical: at batch size 1, cAE–GPU requires 0.1204 s, compared to 0.1221 s for Residual-MLP–GPU; at batch size 1000, the times are 1.0210 s and 1.0493 s, respectively. On CPU, their runtimes also remain close, for example, at batch size 50, cAE–CPU takes 0.6773 s while Residual-MLP–CPU takes 0.5567 s.

Compared to the traditional approximant IMRPhenomXP\_NRTidalv2 running on CPU, both neural models executed on GPU yield speed-ups. For batch size 1, cAE–GPU is approximately 5.54 times faster (0.1204 s vs. 0.6663 s), and for batch size 1000, the speed-up increases to 9.92 times (1.0210 s vs. 10.1279 s). Residual-MLP–GPU shows a similar trend, reaching 9.65 times faster at batch size 1000.

\begin{table*}[htbp]
\centering
\caption{Waveform generation time for different batch sizes. Residual-MLP-CNN is abbreviated as ResMC. ResMC–GPU and ResMC–CPU refer to ResMC evaluated on an A800 GPU and on two Xeon Silver 4214R CPUs, respectively. Similarly, cAE–GPU and cAE–CPU denote our conditional autoencoder model evaluated on GPU and CPU. IMR–CPU refers to IMRPhenomXP\_NRTidalv2 evaluated on the same CPU system. The speed-up ratios are defined as SU(IMR/cAE-G) = t(IMR-CPU) / t(cAE-GPU) and SU(IMR/ResMC-G) = t(IMR-CPU) / t(ResMC-GPU), where t denotes the time.}

\label{tab:batch_efficiency}
\begin{tabular}{cccccccc}
\hline
Batch size & cAE–GPU & cAE–CPU & ResMC–GPU & ResMC–CPU & IMR–CPU & SU(IMR/cAE–G) & SU(IMR/ResMC–G) \\ \hline
1     & 0.1204\,s & 0.2314\,s & 0.1221\,s & 0.1675\,s & 0.6663\,s & 5.54 & 5.46 \\
10    & 0.1500\,s & 0.3235\,s & 0.1318\,s & 0.2797\,s & 0.7113\,s & 4.74 & 5.40 \\
50    & 0.1663\,s & 0.6773\,s & 0.1726\,s & 0.5567\,s & 1.0646\,s & 6.40 & 6.17 \\
100   & 0.2501\,s & 1.2440\,s & 0.2291\,s & 1.1831\,s & 1.1377\,s & 4.55 & 4.97 \\
500   & 0.5902\,s & 5.8146\,s & 0.5927\,s & 5.5440\,s & 5.2263\,s & 8.86 & 8.82 \\
1000  & 1.0210\,s & 11.7420\,s & 1.0493\,s & 11.1276\,s & 10.1279\,s & 9.92 & 9.65 \\
\hline
\end{tabular}
\end{table*}

\paragraph{Bayesian inference efficiency.}  
To further demonstrate the practical utility of our cAE model in downstream applications,  
we performed a simplified Bayesian inference experiment on the binary neutron star masses $(m_1, m_2)$.  
The injected (true) source parameters were set to $m_1^{\rm true} = 1.45\,M_{\odot}$, $m_2^{\rm true} = 1.25\,M_{\odot}$,  
with dimensionless spins $\chi_{1z} = \chi_{2z} = 0.1$ and tidal deformabilities $\Lambda_1 = \Lambda_2 = 200$.  
The detector noise was modeled using the aLIGO O4 sensitivity curve, and posterior sampling was performed with the \texttt{emcee} package\cite{corner}.  
We employed 4 chains of 5000 steps each, discarding the first 1000 steps as burn-in.  
Since the purpose of this experiment is primarily to compare the efficiency of waveform templates,  
rather than to perform full high-dimensional parameter estimation,  
we fixed the spin and tidal parameters and inferred only $(m_1, m_2)$.  
This simplification is sufficient to reflect the computational cost and efficiency differences between traditional approximants and the deep-learning cAE templates.  

\begin{figure}[htbp]
    \centering
    \includegraphics[width=0.45\textwidth]{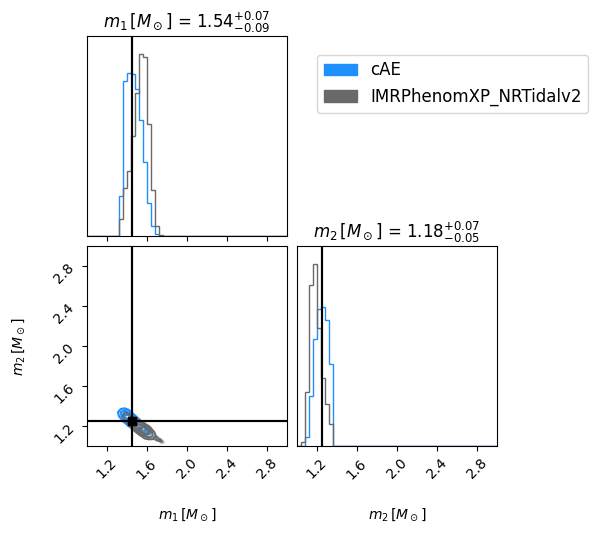}
    \caption{Posterior distributions of binary neutron star component masses $(m_1,m_2)$ 
    obtained from Bayesian inference.}
    \label{fig:corner_plot}
\end{figure}

The inference results are shown in FIG.~\ref{fig:corner_plot}.  
For the cAE template, the estimated parameters are  
$m_1 = 1.469$ with a standard deviation of $0.075$ ($m_1 = 1.469 \pm 0.075\,M_{\odot}$) and  
$m_2 = 1.245$ with a standard deviation of $0.060$ ($m_2 = 1.245 \pm 0.060\,M_{\odot}$),  
with 68\% credible intervals of $(1.387, 1.550)$ and $(1.178, 1.312)$, respectively.  
For the traditional approximant, the estimates are  
$m_1 = 1.530$ with a standard deviation of $0.078$ ($m_1 = 1.530 \pm 0.078\,M_{\odot}$) and  
$m_2 = 1.191$ with a standard deviation of $0.059$ ($m_2 = 1.191 \pm 0.059\,M_{\odot}$),  
with 68\% credible intervals of $(1.427, 1.607)$ and $(1.134, 1.353)$, respectively.  

The results suggest that the cAE model delivers parameter estimates with accuracy on par with the IMRPhenomXP\_NRTidalv2 approximant. In terms of runtime, the cAE model provides a meaningful improvement in computational efficiency. Under the same sampling setup, the traditional approximant required approximately 312 seconds using 24 CPU cores, while the cAE completed the inference in only 61 seconds on a single NVIDIA A800 GPU—resulting in a speedup of approximately 5.1 times. These findings provide initial evidence that the cAE model not only maintains high waveform reconstruction fidelity but also offers significant computational benefits in practical inference workflows. This makes it particularly well-suited for time-sensitive applications such as rapid alerts or large-scale population analyses. We hope this experiment offers a useful reference point for future extensions to more complex, higher-dimensional inference scenarios.

\section{Summary and Discussion}  
\label{sec:Summary and Discussion}
This study presents an efficient gravitational waveform generation method based on a cAE and applies it to amplitude-phase modeling of BNS systems. Compared to traditional waveform approximation methods such as IMRPhenomXP\_NRTidalv2, cAE significantly improves computational efficiency while maintaining high reconstruction accuracy. On a large-scale test dataset, the averaged waveform mismatch is $2.13\times10^{-3}$, corresponding to an average accuracy exceeding $99.79\%$. Even with precession and tidal effects, cAE maintains high precision across different $\Theta$ ranges.  
For efficiency, the cAE model also performs well. On GPU, it takes 0.1204 s to generate a single waveform, while IMRPhenomXP\_NRTidalv2 on CPU takes 0.6663 s, giving a speed-up of about 5.54 times. 
As the batch size increases, this advantage becomes more clear. For batch size 1000, cAE–GPU takes 1.0210 s, while IMRPhenomXP\_NRTidalv2 on CPU takes 10.1279 s, giving a speed-up of about 9.92 times. Residual-MLP–GPU shows similar results across all batch sizes. These results suggest that the cAE model maintains both high precision and relatively higher efficiency in waveform generation, especially when running on GPU.

Although this study has made some progress, several aspects remain worthy of further exploration. Future work can improve waveform reconstruction accuracy and generation efficiency by expanding the training dataset and incorporating more advanced model architectures. Generation speed is also a key factor, particularly critical for real-time gravitational wave applications. We plan to explore inference optimization strategies such as TensorRT \cite{chaturvedi2022inference,lijun2020using} and ONNX Runtime \cite{alizadeh2024green} to further reduce latency and enhance practical applicability. In addition, we will systematically evaluate the impact of latent variable dimensionality on waveform modeling and extend the parameter space to cover more complex physical systems, such as eccentric binaries and sources with stronger spin and tidal effects. On the architectural side, future work may explore diffusion models \cite{Mo2024Scaling}, Transformer variants \cite{fei2024dimba}, and variational autoencoders (VAEs, see FIG.~\ref{fig:vae}) to enhance generation performance and improve generalization.

In conclusion, the cAE-based waveform generation method proposed in this study offers an efficient and accurate approach to BNS waveform modeling. It shows strong potential for real-time data analysis, large-scale parameter estimation, and GW event identification. As deep learning continues to advance, data-driven methods are expected to play a growing role in GW astronomy, providing more precise and computationally efficient tools for signal modeling and fundamental physics research.

\section*{Acknowledgements}

This work was supported by the National Key Research and Development Program of China (Grant No. 2021YFC2203004), the Fundamental Research Funds for the Central Universities Project (Grant No. 2024IAIS- ZD009), the National Natural Science Foundation of China (Grant No. 12575072, Grant No. 12347101), and the Natural Science Foundation of Chongqing (Grant No. CSTB2023NSCQ-MSX0103). The source code of this study is available at \url{https://github.com/thisshy/BNS_Waveform_Generator_Based_on_CAE}.

\bibliography{references}

\clearpage
\appendix

\onecolumngrid

\section{}
\label{sec:appendix}
\vspace{-1em}

\begin{figure*}[!htbp]
    \centering
    \includegraphics[width=0.8\textwidth]{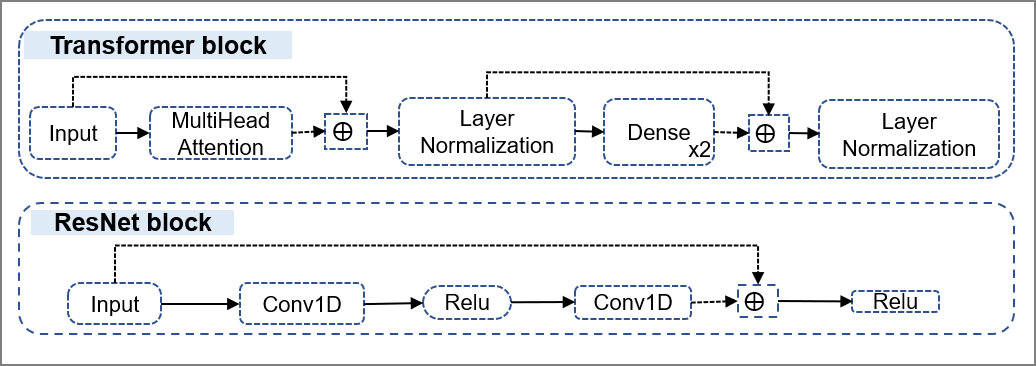}
    \caption{Architectures of the Transformer and ResNet blocks used in our model.}
    \label{fig:resnet_transformer}
\end{figure*}

\begin{figure*}[!htbp]
    \centering
    \includegraphics[width=1\textwidth]{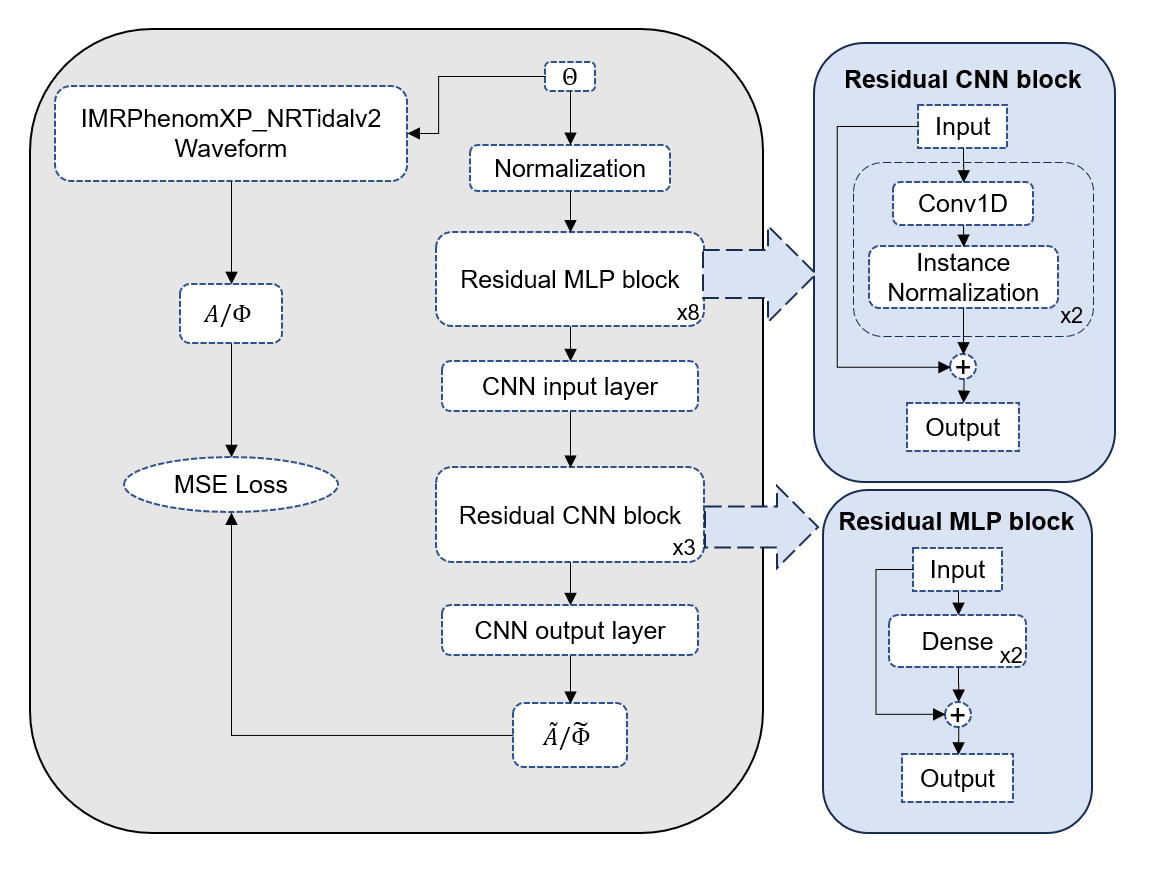}
    \caption{Architectures of the residual-stacked Multilayer Perceptron (MLP) and Convolutional Neural Network (CNN) model.}
    \label{fig:res_mlp_cnn_network}
\end{figure*}

\begin{figure*}[!htbp]
    \centering
    \includegraphics[width=0.8\textwidth]{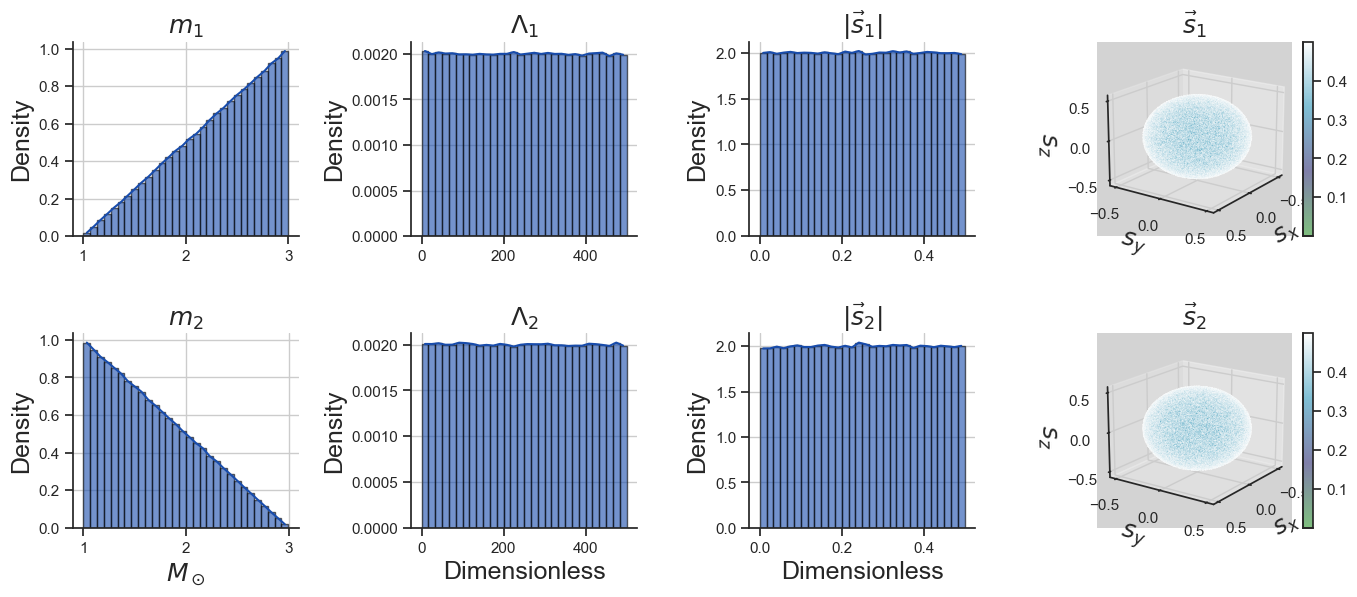}
    \caption{Distribution of the $1\times10^{6}$ training samples in the parameter space $\Theta$, including component masses, tidal deformabilities, and spin components. All parameters are sampled independently using uniform or isotropic priors.}
    \label{fig:training_parameter_distribution}
\end{figure*}

\begin{figure*}[!htbp]
    \centering
    \includegraphics[width=0.8\textwidth]{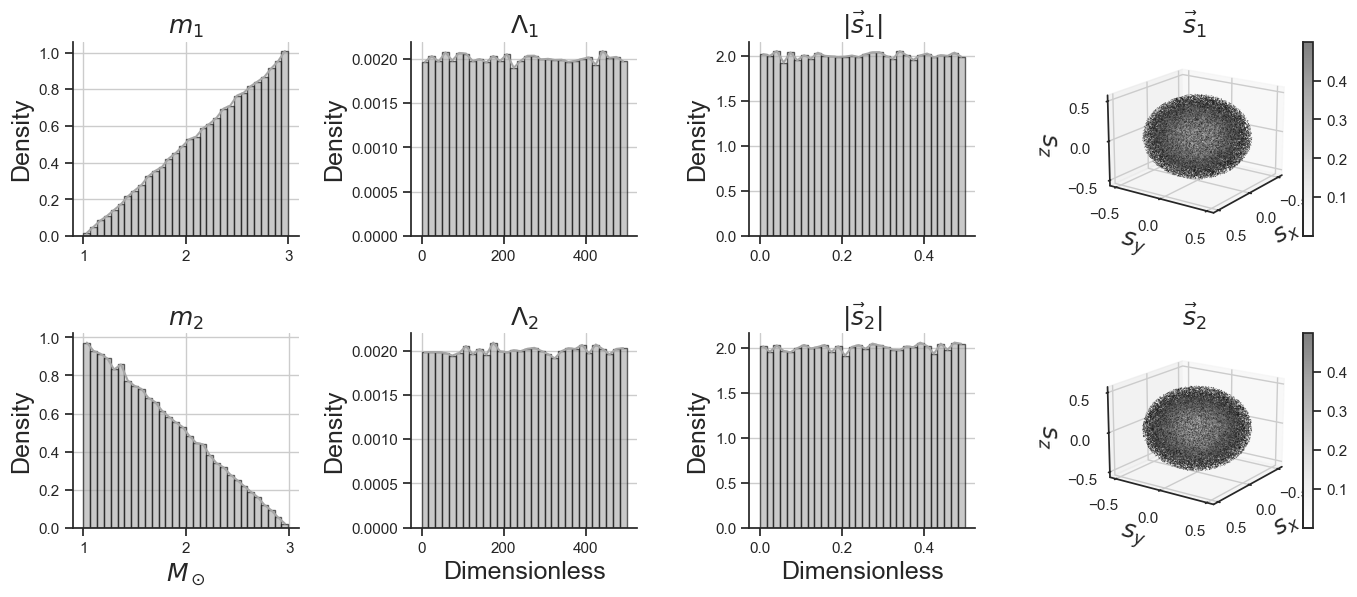}
    \caption{Distribution of the $1\times10^{5}$ testing samples. The parameter priors are identical to those used in training, ensuring a consistent and representative evaluation set.}
    \label{fig:testing_parameter_distribution}
\end{figure*}

\begin{figure*}[!htbp]
    \centering
    \includegraphics[width=0.7\textwidth]{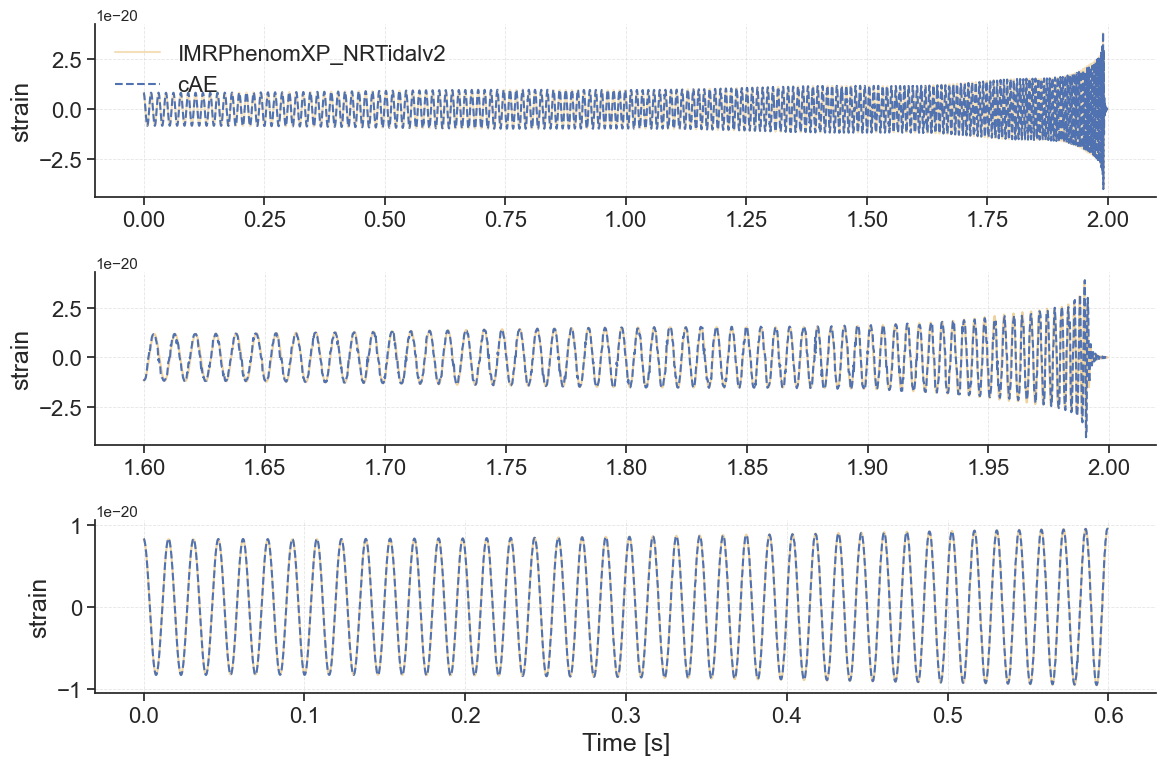}
    \caption{Comparison of gravitational waveforms generated by the cAE (blue) and IMRPhenomXP\_NRTidalv2 (yellow). The top subplot shows the full waveform over $0$–$2$ s, with mismatch = $8.1\times10^{-4}$ and 200  cycles. The middle subplot zooms into $t = 1.5998$–$1.9995$ s. The bottom subplot zooms into $t = 0$–$0.5996$ s. The test waveform parameters are $m_1 = 2.84M_{\odot}$, $m_2 = 2.81M_{\odot}$, $\Lambda_1 = 188.48$, $\Lambda_2 = 233.67$, 
    $\mathrm{spin}_{1x} = 0.13$, $\mathrm{spin}_{1y} = -0.22$, $\mathrm{spin}_{1z} = -0.01$,  
    $\mathrm{spin}_{2x} = -0.09$, $\mathrm{spin}_{2y} = -0.10$, $\mathrm{spin}_{2z} = -0.21$.}
    \label{fig:c_200}
\end{figure*}

\begin{figure*}[!htbp]
    \centering
    \includegraphics[width=0.7\textwidth]{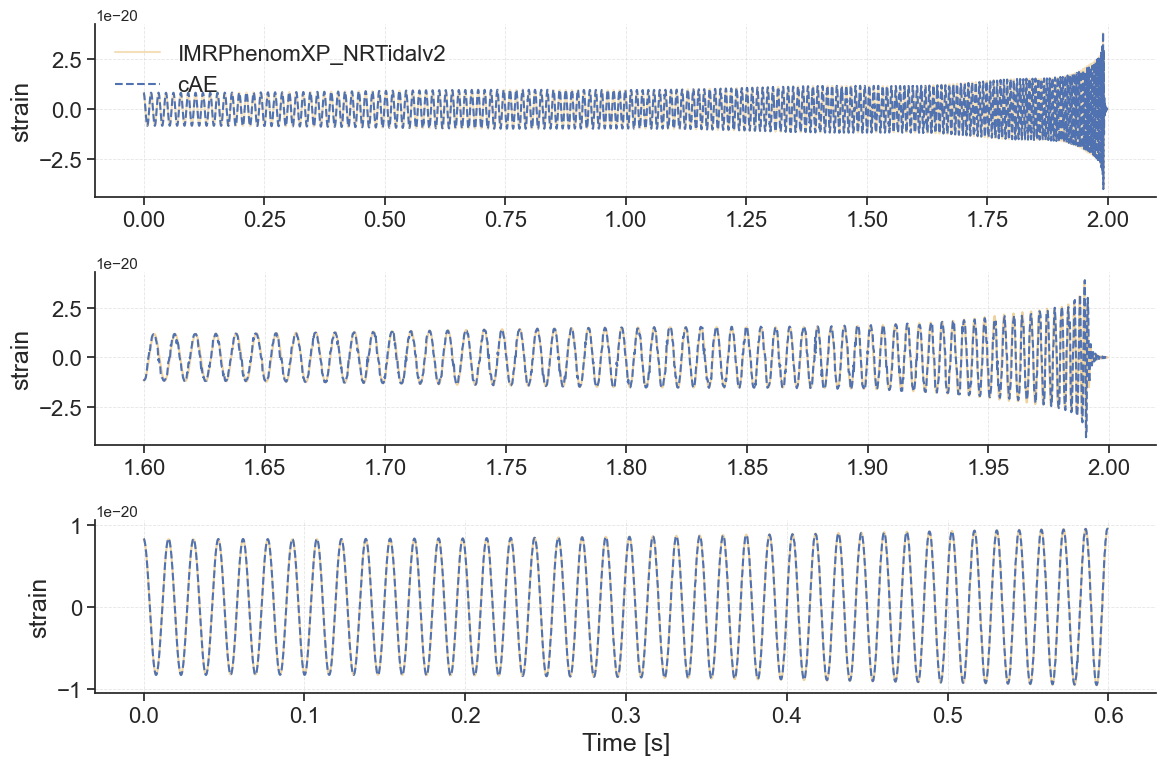}
    \caption{The top subplot shows the full waveform over $0$–$2$ s, with mismatch = $1.2\times10^{-3}$ and 250  cycles. The middle subplot zooms into $t = 1.5998$–$1.9995$ s. The bottom subplot zooms into $t = 0$–$0.5996$ s. The test waveform parameters are $m_1 = 2.66M_{\odot}$, $m_2 = 1.81M_{\odot}$, $\Lambda_1 = 176.83$, $\Lambda_2 = 258.68$,  
    $\mathrm{spin}_{1x} = 0.19$, $\mathrm{spin}_{1y} =0.34$, $\mathrm{spin}_{1z} = 0.08$,  
    $\mathrm{spin}_{2x} = -0.02$, $\mathrm{spin}_{2y} = 0.21$, $\mathrm{spin}_{2z} = -0.18$.}
    \label{fig:c_250}
\end{figure*}

\begin{figure}[h]
    \centering
    \includegraphics[width=0.7\textwidth]{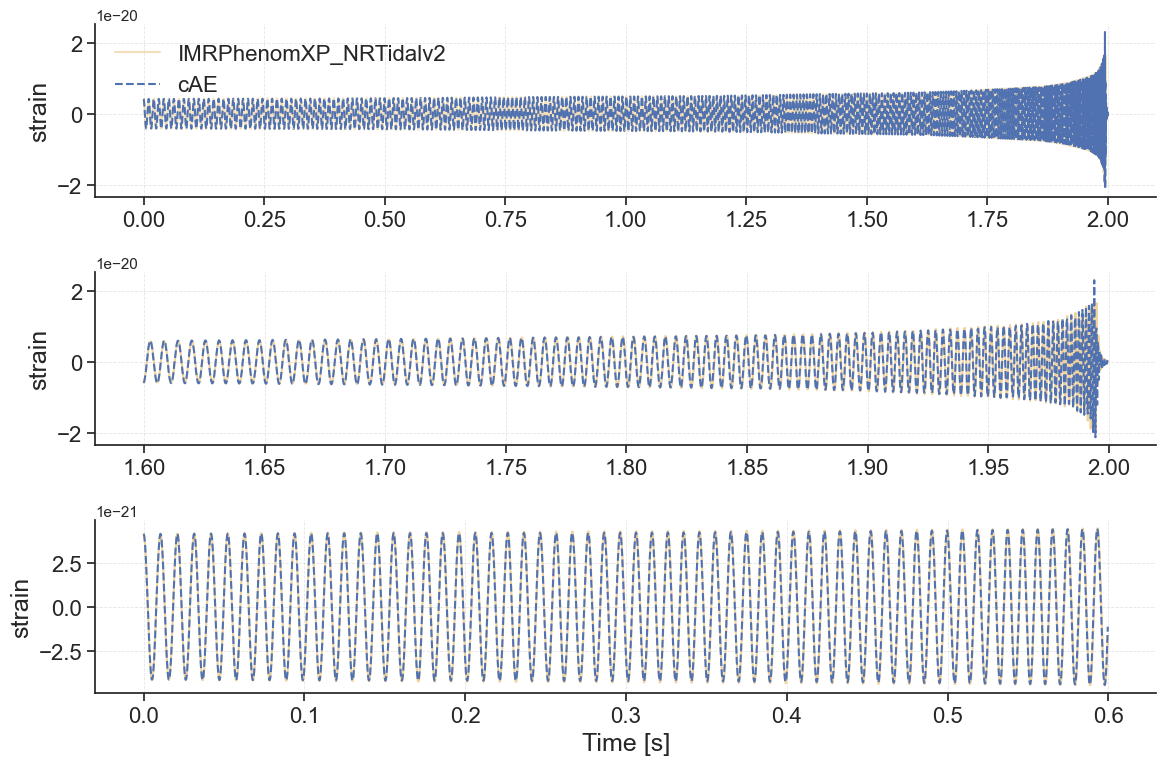}
    \caption{The top subplot shows the full waveform over $0$–$2$ s, with mismatch = $3.4\times10^{-3}$ and 300 cycles. The middle subplot zooms into $t = 1.5998$–$1.9995$ s. The bottom subplot zooms into $t = 0$–$0.5996$ s. The test waveform parameters are $m_1 = 1.69M_{\odot}$, $m_2 = 1.66M_{\odot}$, $\Lambda_1 = 179.26$, $\Lambda_2 = 375.52$,  
    $\mathrm{spin}_{1x} = -0.08$, $\mathrm{spin}_{1y} = 0.20$, $\mathrm{spin}_{1z} = -0.36$,  
    $\mathrm{spin}_{2x} = 0.02$, $\mathrm{spin}_{2y} = 0.36$, $\mathrm{spin}_{2z} = -0.04$.}
    \label{fig:c_300}
\end{figure}

\begin{figure}[h]
    \centering
    \includegraphics[width=0.7\textwidth]{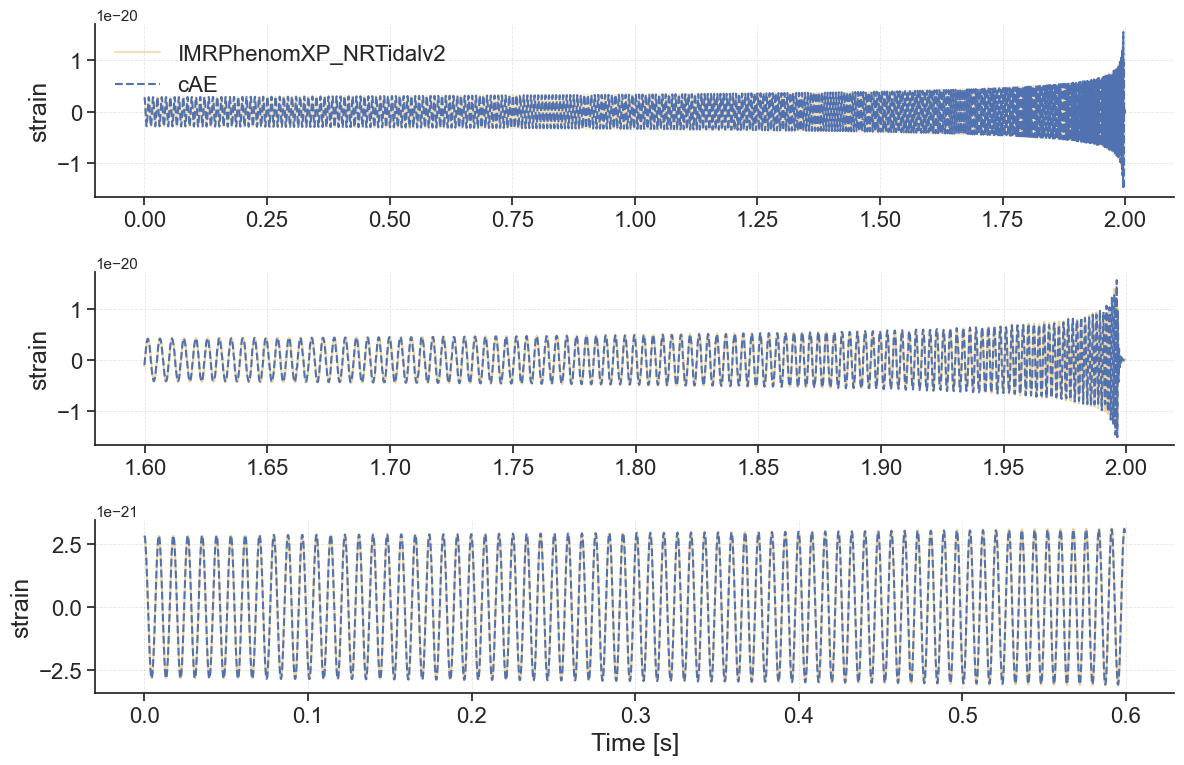}
    \caption{The top subplot shows the full waveform over $0$–$2$ s, with mismatch = $5.7\times10^{-3}$ and 350  cycles. The middle subplot zooms into $t = 1.5998$–$1.9995$ s. The bottom subplot zooms into $t = 0$–$0.5996$ s. The test waveform parameters are $m_1 = 1.31M_{\odot}$, $m_2 = 1.24M_{\odot}$, $\Lambda_1 = 135.05$, $\Lambda_2 = 417.58$,  
    $\mathrm{spin}_{1x} = 0.24$, $\mathrm{spin}_{1y} = 0.03$, $\mathrm{spin}_{1z} = 0.06$,  
    $\mathrm{spin}_{2x} = -0.03$, $\mathrm{spin}_{2y} = -0.30$, $\mathrm{spin}_{2z} =0.32$.}
    \label{fig:c_350}
\end{figure}

\end{document}